\documentclass[aps,prl,preprint,groupedaddress,superscriptaddress,amsmath,amssymb]{revtex4-1}[10pts]
\usepackage{amsmath}
\usepackage{amssymb}
\usepackage{multirow}
\usepackage{graphicx}  
\usepackage{dcolumn}   
\usepackage{bm}        
\usepackage{verbatim}   
\usepackage{lipsum}
\usepackage{graphicx}
\usepackage{braket} 
\usepackage{ulem}
\usepackage{color}

\begin{document}

\title{Visualizing the Phase-Space Dynamics of an External Cavity Semiconductor Laser}
\author{C.~Y.~Chang}
\email{cychang@gatech.edu} 
\affiliation{Georgia Institute of Technology, School of Physics, Atlanta, Georgia 30332-0250 USA}%
\affiliation{%
UMI 2958 Georgia Tech-CNRS, Georgia Tech Lorraine, 2 Rue Marconi F-57070, Metz, France}%

\author{Michael J. Wishon}
\affiliation{%
	UMI 2958 Georgia Tech-CNRS, Georgia Tech Lorraine, 2 Rue Marconi F-57070, Metz, France}%
\affiliation{School of Electrical and Computer Engineering, 
Georgia Institute of Technology, Atlanta, Georgia 30332-0250 USA}%

\author{Daeyoung Choi}%
\affiliation{%
	UMI 2958 Georgia Tech-CNRS, Georgia Tech Lorraine, 2 Rue Marconi F-57070, Metz, France}%
\affiliation{School of Electrical and Computer Engineering, 
Georgia Institute of Technology, Atlanta, Georgia 30332-0250 USA}%

\author{K. Merghem}
\author{Abderrahim Ramdane}
\affiliation{%
Center for Nanosciences and Nanotechnologies (CNRS-C2N), Route de Nozay, 91460 Marcoussis-France}%

\author{Fran\c{c}ois Lelarge}
\email{Present address: Almae Technologies, Route de Nozay, 91460 Marcoussis- France}
\affiliation{%
Center for Nanosciences and Nanotechnologies (CNRS-C2N), Route de Nozay, 91460 Marcoussis-France}%


\author{A. Martinez}
\affiliation{%
Center for Nanosciences and Nanotechnologies (CNRS-C2N), Route de Nozay, 91460 Marcoussis-France}%

\author{A. Locquet}
\affiliation{%
	UMI 2958 Georgia Tech-CNRS, Georgia Tech Lorraine, 2 Rue Marconi F-57070, Metz, France}%
\affiliation{School of Electrical and Computer Engineering, Georgia Institute of Technology, 
Atlanta, Georgia 30332-0250 USA}%

\author{D. S. Citrin}
\email{david.citrin@ece.gatech.edu}
\affiliation{%
	UMI 2958 Georgia Tech-CNRS, Georgia Tech Lorraine, 2 Rue Marconi F-57070, Metz, France}%
\affiliation{School of Electrical and Computer Engineering, 
Georgia Institute of Technology, Atlanta, Georgia 30332-0250 USA}%

\date{\today}
\begin{abstract}
We map the phase-space trajectories of an external-cavity semiconductor laser using phase portraits.
This is both a visualization tool as well as a thoroughly quantitative approach enabling
unprecedented insight into the dynamical regimes, 
from continuous-wave through coherence collapse as feedback is increased.
Namely, the phase portraits in the intensity \textit{versus} laser-diode terminal-voltage 
(serving as a surrogate for 
inversion) plane are mapped out. 
We observe a route to chaos interrupted by two types of limit cycles, 
a subharmonic regime and period-doubled dynamics at the edge of chaos. 
The transition of the dynamics are analyzed utilizing bifurcation diagrams for 
both the optical intensity and the laser-diode terminal voltage. 
These observations provide visual insight into the dynamics in these systems.

\end{abstract}

\maketitle
\section{Introduction} 	

The complex dynamics experienced by a nonlinear system for given initial conditions corresponds to 
a trajectory in phase space that span all the system's dynamical variables. 
Visualizing the phase-space dynamics both provides a heuristic tool as well as
opens a quantitative window onto the dynamics that may otherwise be difficult to apprehend.
For a laser diode (LD) with time-delayed optical feedback, 
otherwise known as an external-cavity semiconductor laser (ECL), 
a three dimensional projection is typically used. This projection is spanned by the optical intensity$I\!\propto \! \left| E \right| ^2$ (with $E$ the
slowly-varying amplitude of the optical field), the optical phase $\phi$, and the inversion $n$ 
in the gain medium. The high speed and high dynamical dimensionality of ECLs has attracted broad 
interest for various applications, ranging from radar \cite{LinJQE}, 
ultrafast random bit generation \cite{UchidaNatPho,NianqiangOE}, and optoelectronic 
oscillators \cite{ChangAPL,Chang2017JQE}, to name a few. 
Moreover, the intrinsic dynamics of ECLs have attracted attention to implement a 
number of information-processing functionalities in the physical layer, 
including chaotic encrypted data communications \cite{ArgyrisNature,DamienPRE,ZhangOL}, 
compressive sensing \cite{DamienSR}, and reservoir computing \cite{LargerPRX,BuenoOE}. 
But to realize the full potential of such applications, a detailed knowledge of the complex 
dynamics of ECLs is required \cite{DaanPR}, and despite 50 years of study, there are basic issues that remain to be elucidated. In Ref. \cite{BrunnerPRL}, phase plots of the optical intensity, phase, and laser voltage were obtained but only for the complex chaotic regime of low-frequency fluctuations. In this paper, we focus on phase portraits revealing the first instabilities that lead the ECL into a chaotic regime. 

It has been pointed out, based on delay embedding theory, that the phase portrait in $|E|$ and $n$ can be reconstructed by plotting $|E(t)|$ versus $|E(t-\tau)|$ parametrically in time.  
Refs. \cite{Taken} and \cite{Sauer} investigate the requirements for accessing this information by reconstructing from the signal and delayed signal in other deterministic time-delayed systems. Theoretically, the reconstructed phase portraits and the experimental phase portraits may share some (invariant) properties; however, they are limited to the vicinity of an attractor along with other mathematical conditions discussed in Ref. \cite{whitney}.   Note, however, it is not a foregone conclusion that such reconstructed phase portraits are necessarily correct, even proximate to an attractor.  Clearly, the ability to obtain \textit{experimental} phase portraits takes precedence over any reconstructed phase portraits.  Our work opens the way for such a detailed comparison, though the aim of the present work is to introduce the technique rather than to provide such a comparison.

From a more general perspective, ECLs provide time-delayed feedback systems in which a number of key parameters can be controlled, and combining this with their relative ease of implementation on the tabletop
or within photonics integrated circuits, 
these systems serve as excellent test-beds for the bifurcation analysis for high-dimensional 
nonlinear systems \cite{KBerndAIP,NianqiangJOE}. The dynamics of ECLs are commonly 
described by the Lang-Kobayashi (LK) model \citep{LangJQE}. 
The LK model provides a coupled delay-differential-equation description of the system 
variables, \textit{viz.}\ $E$, $n$, and $\phi$.
By adjusting various control parameters (feedback strength $\eta$, external-cavity length $L\!=\!c/(2\tau)$
with $\tau$ the feedback time, injection current $J$), 
various dynamical regimes can be 
accessed \cite{tromborgPTL,kao1994IEEE,ahmed2009numerical,MukaiPRL}. 
  
The LK model allows one to explore dynamical trends as various parameters are tuned; 
most studies focus on $\eta$ \cite{KimOE,MorkPRL,HohlPRL,LiJQE} and visualize these 
trends by means of bifurcation diagrams (BD). Recently, we 
demonstrated experimental BDs for the optical intensity $I(t)$ 
as $\eta$ is varied \cite{KimJQE,KimPRA,tromborgPTL}. 
Both theoretical and experimental BDs provide global perspectives on the evolution of the 
dynamical regimes as the chosen parameter is varied; 
in particular, one can test predictions of the sequence of bifurcations based on the LK model. 
Most studies to date, however, only focus on $I(t)\!\propto\!E^2(t)$ \cite{note1}. 
To more thoroughly map the dynamics in phase space, it is necessary to probe additional 
dynamical degrees of freedom. Changes in $n$ in the gain region can be obtained from 
the LD terminal voltage $V(t)$ at fixed $J$ \cite{kazarinov, RayPRE,14sahai}. 
We recently reported such measurement \cite{note}, while varying the control parameter $\eta$, in 
Refs.\  \cite{ChangAPL,Chang2017JQE}. This allowed us to plot BDs based on $V(t)$. 
These investigations provide another component of a global perspective of the route to chaos in ECLs.
In Ref.\  \cite{Chang2017JQE}, a largest-Lyapunov-exponent analysis was applied to 
both $I(t)$ and $V(t)$ verifying that the measurement of $V(t)$ provides information 
of comparable dynamical complexity to that provided by $I(t)$. 
In order to interpret the information contained in this additional degree of freedom, 
it is essential to focus on its dynamics compared to the better-known dynamics in $I(t)$. 
While their relationship has been studied 
theoretically \cite{DaanPR,KBerndAIP,MasollerPRA,BrunnerPRL} in various dynamical regimes, 
to date experimental verification, outside of LFF, is absent. 

In this study, we present \textit{simultaneous measurements of the time series (TS) $I(t)$ and $V(t)$
on the route to chaos, from continuous-wave (CW) to coherence collapse (CC)}. 
The TSs, when plotted parametrically in time $t$ in the $I$-$V$ plane 
($V$ serving as a surrogate for $n$) provide a \textit{phase portrait} of the dynamics, \textit{i.e.}, 
the resulting plot is a projection in the $\phi$ direction of the phase-space trajectory onto the $I$-$V$ plane, where $\phi$
is the phase difference between the input and output electric fields.
While numerous \textit{theoretical} studies of phase portraits in ECLs based on the 
LK model have been presented in the literature, \textit{e.g.}, \cite{wieczorek2003bifurcation, DaanPR}, 
to our knowledge, the \textit{experimental} phase portraits presented here are the first published
mapping out the route to chaos in an ECL. 
Together with the global view provided by the BDs \cite{ChangAPL,KimOE,KimJQE,KimPRA,NianqiangOE}, 
the phase portraits provide 
unprecedented detailed insight into the phase-space dynamics in  various  regimes and
open the way to more thorough testing of theoretical predictions.

Before embarking on our more detailed discussion, it is helpful to orient ourselves by  means of simple examples.
To start, consider an ECL  under CW operation, for  which   
$I(t)$ and $V(t)$ are constant. 
In this case, a parametric plot of $I(t)$ and $V(t)$ in $t$ will be a point in the $I$-$V$ plane. 
Now consider relaxation oscillations (RO), a small-signals model accounting 
for the dynamics of $I(t)$ and $n(t)$ given a small temporal perturbation shows 
that $I(t)$ and $n(t)$ both exhibit damped sinusoidal oscillations near the LD 
RO frequency $f_{RO}$ with $n(t)$ lagging $I(t)$ by $\pi/2$ \cite{Yariv}. 
In an ECL, for certain ranges of $\eta$, these ROs may become undamped and shifted in 
frequency \cite{tromb,cohen,schunk,helms,RitterJOSA,van}. 
Consequently, a parametric plot of $I(t)$ and $V(t)$ in $t$, again the corresponding phase portrait, 
will be an ellipse with semi-major axis with negative 
slope in the $I$-$V$ plane. Of course, in other dynamical regimes, 
the phase portrait will reflect the more complex combined dynamics of $I(t)$ and $V(t)$, as we see  below.

\section{Experimental Setup}

\begin{figure}[t]

\centerline{\includegraphics[width=.8\textwidth]{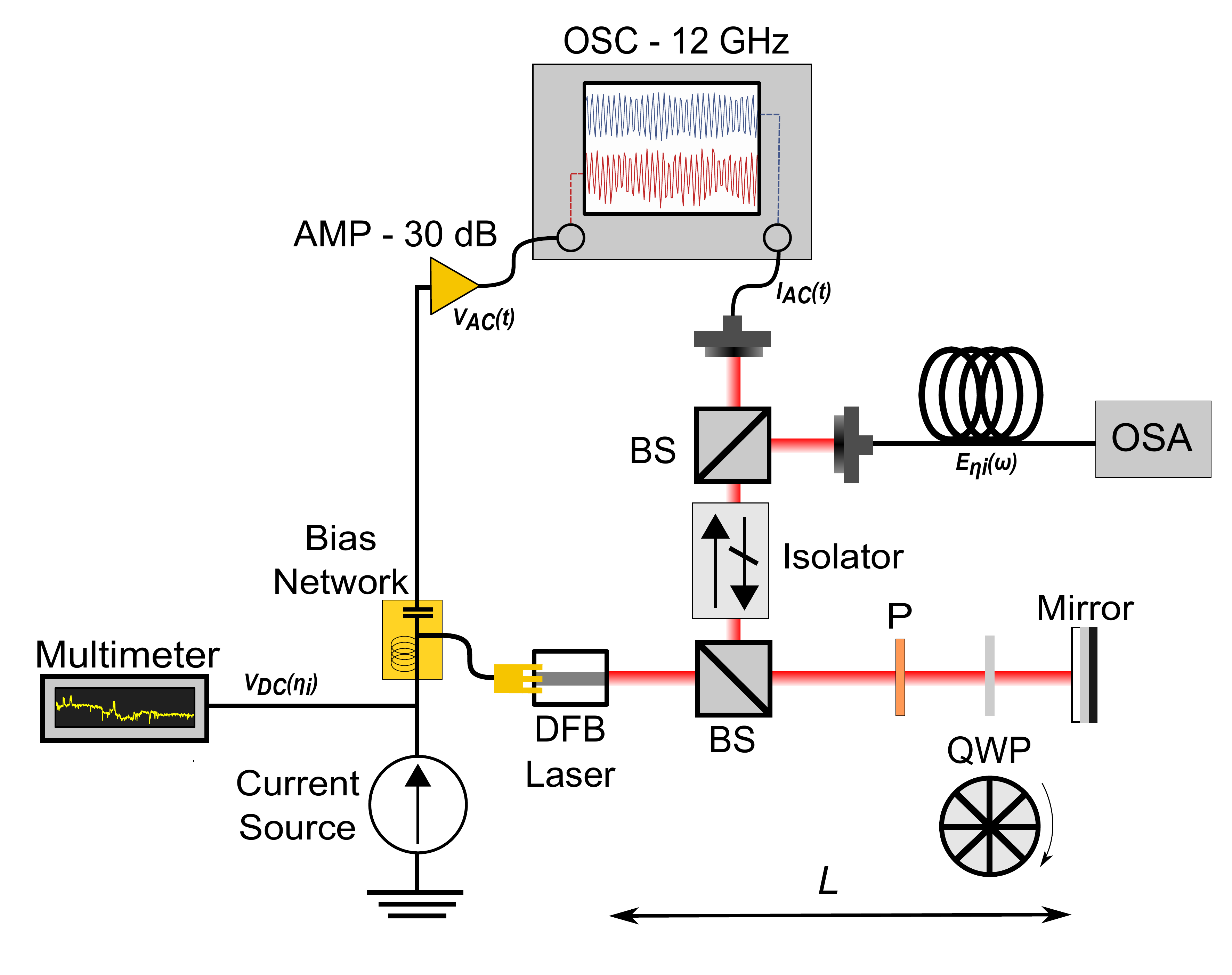}}
\vspace{-5pt}
\caption{Experimental setup. OSC: oscilloscope, AMP: radio-frequency (RF) amplifier, OSA: 
optical spectrum analyzer, BS: beam splitter, P: polarizer, QWP: quarter-wave plate.}
\vspace{-10pt}
\label{figure1}
\end{figure}

A schematic of our ECL setup is shown in Fig.\ \ref{figure1}. For our experiments, we 
employ a single longitudinal-mode edge-emitting InGaAsP distributed feedback (DFB) 
LD containing seven quantum wells in the active region \cite{ChangAPL}. The grating 
is designed and fabricated to achieve a $k$ product of 50 cm$^{-1}$ and the length $l$
of the LD cavity is 600 $\mu$m, achieving a $kl$ value of 3. 
The LD is that of Ref. \cite{AnthonyAPL}. The LD emits at 1550 nm 
with a free-running threshold current $J_{th}\!=\!29.8$ mA. 
We simultaneously measure $I(t)$ and $V(t)$
where we synchronize the two via cross-correlation of two signals as the optical path length is varied;
the signals are synchronized to within $1$ ps, which is much shorter than any relevant 
dynamical timescale.

During the experiment, the OSA continuously measures the optical spectrum 
for each $\eta$. $V(t)$ is measured across the LD injection terminals utilizing a radio-frequency (RF) probe 
(Cascade Microtech AE-ACP40-GSG-400) with 40-GHz bandwidth. The DC and AC components of 
$V(t)\!=\!V_{DC}+V_{AC}(t)$ are separated with a bias tee (Keysight 11612A). 
The AC component, $V_{AC}(t)$, is amplified (AMP) with an 30-dB amplifier (Microsemi UA030VM) 
with 30 GHz bandwidth in real-time. In addition, the AC components $I_{AC}(t)$ [of $I(t)=
I_{DC}+I_{AC}(t)$] and $V_{AC}(t)$ 
are simultaneously recorded on a real-time oscilloscope (OSC) (Agilent DSO80804B) 
with 12 GHz cut-off frequency. 
The AC signals are small compared with the DC signals leading to the approximate
proportionality of these signals to the dynamical relevant degrees of 
freedom $E(t)$ and $n(t)$ \cite{kazarinov}.

To tune $\eta$, the relative angle between the polarizer (P) and quarter-wave plate (QWP) in 
Fig.\ 1 is adjusted in $0.05^{\circ}$ increments ($\eta \!= \!1$ corresponds to $\sim$15 \% 
of the optical power coupled back onto the collimating lens). 
We intentionally reduce the feedback coupling (estimated one-third of the feedback 
power arrives at the active region of the LD compared to our previous work \cite{ChangAPL,Chang2017JQE}). 
This is done by reducing the overlap of the optical mode of the feedback field into the active region of the LD in order to extract a clear BD. 

\section{Results and Discussion}

As we increase $\eta$, $E(t)$ and $E(t-\tau)$
interact within the nonlinear active region of the LD, 
giving rise to the various types of dynamics discussed below. These dynamics are manifested in 
both $n(t)$ [measured through $V(t)$] and $E(t)$ [measured through $I(t)$]. 
The TSs at the various values of $\eta$ can be used to plot the BDs shown in this section. 
We start with a 
global picture of the route to chaos presented in terms of BDs for $I(t)$ and $V(t)$ as $\eta$ is increased. 
In the subsections below, we 
consider  phase portraits within various dynamical regimes along the route to chaos.

\begin{figure}[t]
\vspace{-10pt}
\centerline{\includegraphics[width=.6\textwidth]{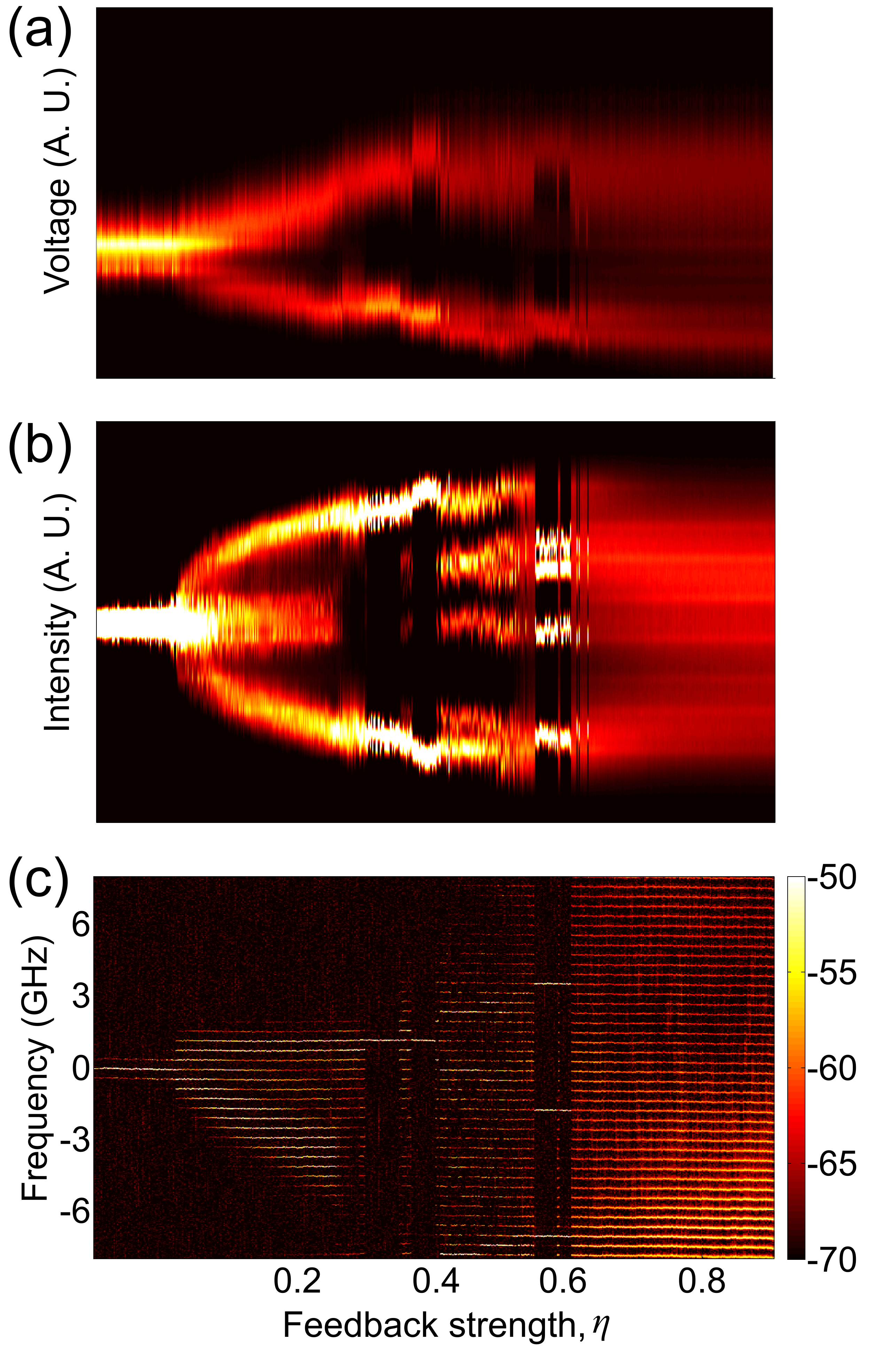} }
\vspace{-5pt}
\caption{BDs illustrating the various dynamical regimes as $\eta$ is varied: (a) for $V(t)$ and (b)
for $I(t)$.
In (b), various dynamical regimes, listed in Table I, are labelled. (c) The optical spectrum is 
taken simultaneously showing the active ECMs as $\eta$ is varied;
the zero of the frequency scale corresponds to the LD free-running frequency ($\eta\!=\!0$)
for the value of $J$ used. All data are taken for $J \!= \!90$ mA and $L \!= \!42$ cm.
Color scales provide relative values of the density but are in arbitrary units.}
\label{bd2}
\vspace{-10pt}
\end{figure}

Figure \ref{bd2} shows the BD obtained from (a) $V(t)$ and (b) $I(t)$ as a function of $\eta$. 
Each BD is obtained by plotting the density of local extrema of the corresponding TS 
as a function of $\eta$. The density is high in white, intermediate in shades of orange, but low in black.
The value  reflects the turning point of the trajectories 
(maxima and minima in both $V$ and $I$), and thus for a given $\eta$, 
regions showing small ranges of pronounced high- and 
low-density contrast tend to represent dynamical regimes exhibiting 
repetitive (more stable) trajectories. 
The region for $\eta\!>\!0.64$  in Fig.\!  \ref{bd2}(b), however, does not correspond to a small number of
maximum/minimum values of $I(t)$, but to chaotic behavior.  [Note that the BD for $V(t)$ in
Fig.\  \ref{bd2}(a) shows less clear features due to additional noise and filtering in the 
voltage measurement; nonetheless, 
as noted above
we have confirmed that $V(t)$ contains 
dynamical complexity comparable 
to that of $I(t)$ indicating that any filtering of $V(t)$ retains information on crucial information-theroretic 
timescales.]
In Fig.\ \ref{bd2}(b) the dynamical regime is labelled with the 
abbreviation of each regime \cite{KimPRA}; see Table I for the ranges of $\eta$
and their correspondence to dynamical regime.

\begin{table}
\caption{\label{tab:5/tc} Dynamical regimes and the ranges of $\eta$ shown in Fig.\ \ref{bd2}(b).}
\begin{ruledtabular}
\begin{tabular}{ cc } 
Dynamical regime & $\eta$ Range \;\\
\hline
Continuous-wave (CW)&  $[0,0.07]$ \\ 
Quasi-periodic (QP) & $(0.07,0.29]$ \\ 
Limit-cycle (LC) & $(0.29,0.4]$ \\ 
Subharmonic (SH) & $(0.4,0.52]$ \\ 
Period-doubled (PD) & $(0.52,0.64]$ \\ 
Coherence-collapse (CC) & $(0.64,0.9]$  \\ 
\end{tabular}
\end{ruledtabular}
\end{table}

The system evolution can be interpreted as trajectories in phase space in the vicinity of external-cavity modes (ECMs) that are the stable steady-state solutions of the LK equations;
the unstable solutions are known as antimodes \cite{LangJQE}.  
Each ECM is associated  with a different standing-wave solution in the external cavity, 
and thus the various ECMs are separated in frequency approximately by the external-cavity free-spectral range $f_{\tau}\!=\!\tau^{-1}\!=\!c/(2L) = 0.36$ GHz for $L=42$ cm. We can track the ECMs that participate in the dynamics for a given $\eta$ 
from the optical spectrum as in Fig.\! \ref{bd2}(c).
The bright features in Fig.\! \ref{bd2}(c) at each $\eta$ show the active ECMs.  
For example, we can identify the dominant ECM in the periodic regime ($0.29\! \le \!\eta\!\le \!0.4$); 
it is the third ECM  (\textit{i.e.}, ECM 3) above the minimum-linewidth mode (MLM), 
labeled ECM 0, as shown in Fig.\ \ref{bd2}(c).  
The optical spectrum in this regime shows, in addition, many active ECMs $\lesssim\!3$.  
 In the subsections below, we explore the phase portraits as $\eta$ increases taking the ECL from CW to CC.

\subsubsection{Continuous-wave and quasi-periodic}

\begin{figure}[h]
\vspace{-5pt}
\centerline{\includegraphics[width=.7\textwidth]{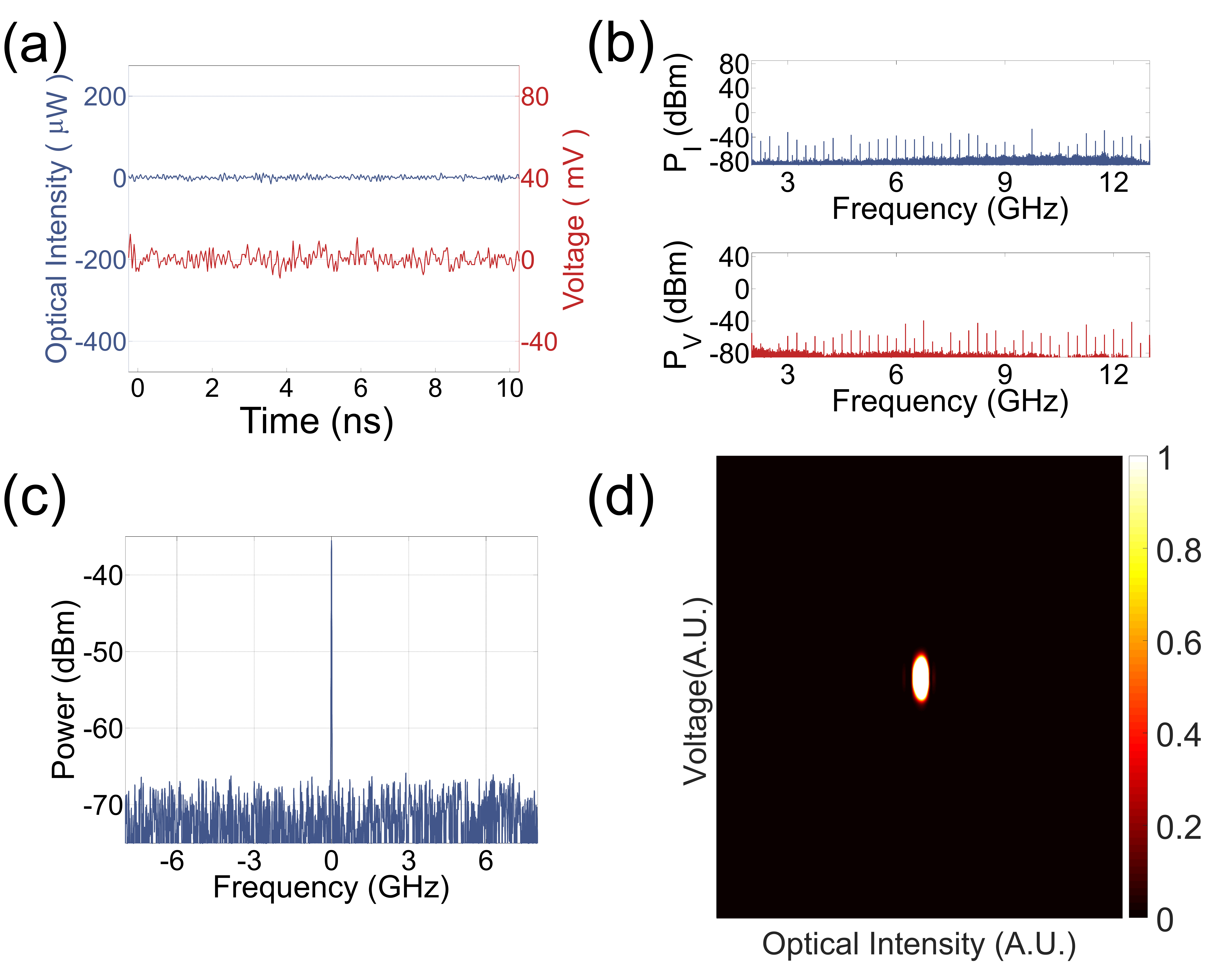}}
\vspace{-5pt}
\caption{The dynamics in the CW regime with $\eta \! = \!0$. 
(a) $I_{AC}(t)$ and $V_{AC}(t)$ and (b) corresponding RF spectra.
 (c) The optical spectrum  and (d)  phase portrait.}
\vspace{-5pt}
\label{CW1}
\end{figure}

We begin with no feedback $\eta\!=\!0$ with the ECL operating in CW. 
CW behavior persists as $\eta$ is increased up to 0.07
as can be seen in Fig.\! \ref{bd2}(b). (See also Table I.)
Figure \ref{CW1}(a) shows both $I(t)$ and $V(t)$ for $\eta\!=\!0$; $V(t)$ and $I(t)$ are constant with 
superimposed noise.  The phase portrait, in Fig.\! \ref{CW1}(d),
consequently, is a dot that is broadened in both the $I$ and $V$ directions by noise. 
Under these conditions, the ECL operates purely on the the free-running LD
frequency, \textit{i.e.}, ECM 0, the minimum-linewidth mode
(MLM) at  wavelength $1550$ nm. [We reference all optical frequencies to $f_{MLM}$ in the plots below. 
In other words, the MLM frequency $f_{MLM}$ in the figures 
corresponds to the zero of the optical-frequency scale.]

The Fourier transform of the TS is commonly referred to as the radio-frequency (RF) spectrum
to distinguish it from the optic spectrum, \textit{i.e.}, the Fourier transform of the optical field.
The RF spectra of $I(t)$ and $V(t)$ are shown in Fig.\! \ref{CW1}(b). Note that they
are broad and featureless, reflecting the near-constant state of $I(t)$ and $V(t)$
with broadband noise under CW operation. Figure \ref{CW1}(c) shows the optical spectrum obtained at
the same time as the TSs. Note the single dominant peak at the MLM, ECM 0.

\begin{figure}[!h]
\vspace{-1pt}
\centerline{\includegraphics[width=.7\textwidth]{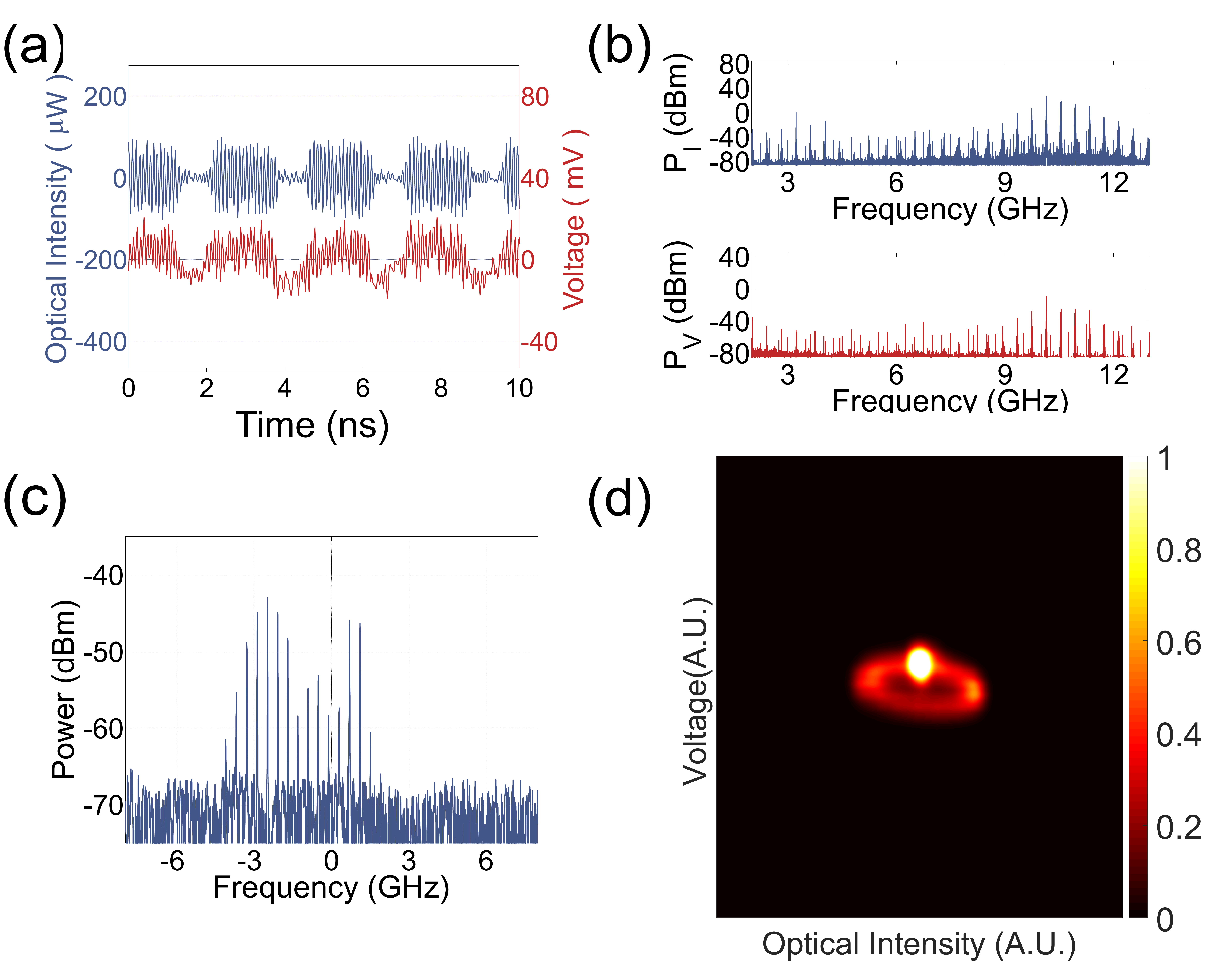}}
\vspace{-5pt}
\caption{The dynamics in the QP regime with $\eta\! = \!0.16$. 
(a) $I_{AC}(t)$ and $V_{AC}(t)$  and (b) corresponding RF spectra.
 (c) The optical spectrum  and (d)  phase portrait.}
\vspace{-10pt}
\label{qppd}
\end{figure}

As $\eta$ is increased above 0.07, but remaining less than 0.29, the dynamical behavior 
undergoes a qualitative
change, known as a Hopf bifurcation, to quasi-periodic (QP). 
 QP denotes dynamics 
having features that appear periodically, though strictly speaking, the TS is not periodic,
as seen in Fig.\ \ref{qppd}(a). 
The RF frequencies contributing to the TSs are concentrated above $\sim$8 GHz, shown in Fig.\ \ref{qppd}(b). 
Many active ECMs ($\sim-10$ to 4) participate, as seen in Fig.~\ref{qppd}(c). 

The observed QP dynamics can be described as a periodic switching between
CW (constant) and limit-cycle (LC) (oscillatory) behaviors (a LC is a closed trajectory in phase space);
 the LC-like behavior which originates from ROs, becomes more dominant 
as $\eta$ increases across this range with the intervals of CW-like behavior 
becoming relatively shorter. In Fig.\! \ref{qppd}(d) we see the corresponding phase portrait.  There
is a central broadened bright spot corresponding to the CW-like 
behavior and a ring-shaped feature corresponding to the LC-like behavior.

\subsubsection{Multiple limit cycles}

\begin{figure}[t]
\centerline{\includegraphics[width=.4\textwidth]{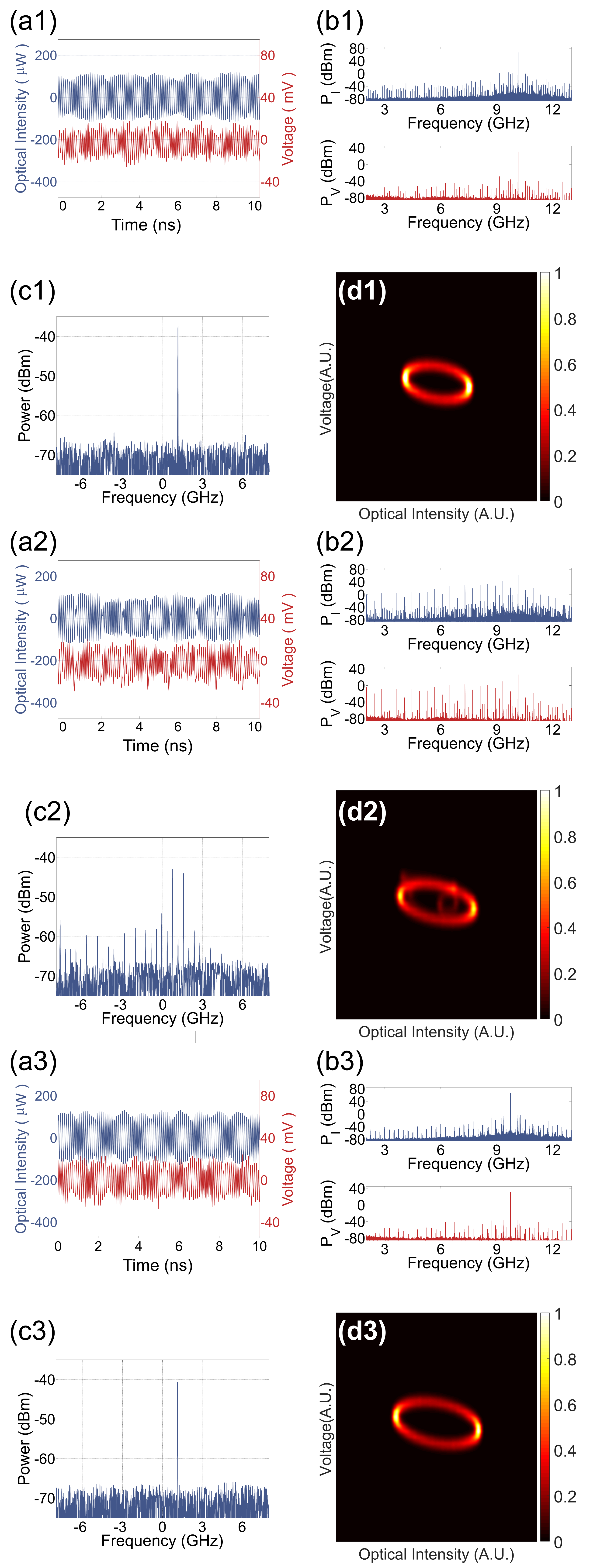}}
\caption{ The dynamics in the LC regime with 
(a1), (b1), (c1), (d1) $\eta\! = \!0.303$, 
(a2), (b2), (c2), (d2) $\eta\! = \!0.365$, and 
(a3), (b3), (c3), (d3) $\eta\! = \!0.383$.}
\vspace{-10pt}
\label{per_all}
\end{figure}

At $\eta\!=\!0.29$, the ECL undergoes a bifurcation into a LC regime \cite{RitterJOSA,RitterJQE};
at this point, CW-like behavior no longer appears. Several LCs are observed for various $\eta$ in this regime,
as seen in the region labelled LC in Fig.\! \ref{bd2}(b) where, apart from narrow ranges of $\eta$, the BD exhibits two high-density features for a given $\eta$
consistent with the approximately sinusoidal behavior of the corresponding TSs as discussed below.
Two such LCs are shown in Fig.\! \ref{per_all}(a1) (TSs), (b1) (RF spectra), (c1) (optical spectrum), and (d1) (phase portrait) and in Fig.\! \ref{per_all}(a3), (b3), (c3), and (d3). [Figure \ref{per_all}(a2), (b2), (c2), and (d2), however, show the corresponding plots
in a narrow QP region between the two distinct LCs, as discussed shortly.] 
These closed trajectories correspond to RO-like oscillations that become undamped in the 
ECL \cite{cohen} in the coupled dynamics
of $I(t)$ and $V(t)$.  

We begin with the LC behavior shown in Fig.\! \ref{per_all}(a1), (b1), (c1), and (d1)
and in Fig.\! \ref{per_all}(a3), (b3), (c3), and (d3). The TSs in Fig.\! \ref{per_all}(a1) and (a3)
show a strong sinusoidal modulation near the RO frequency with a more slowly modulated envelope.
The corresponding RF spectra in Fig.\! \ref{per_all}(b1) for $\eta\!=\!0.303$ is $f_{RO}$, while that in
Fig.\! \ref{per_all}(b3) for $\eta\!=\!0.383$ is $f_{RO}\!-\!f_{\tau}$.  
The optical spectra Fig.\! \ref{per_all}(c1) and (c3) show a single
dominant ECM 3 for these two LCs.  Lastly, the phase portraits Fig.\! \ref{per_all}(d1) and (d3)
exhibit an elliptical structure with a slope indicating a $\pi/2$ phase,
consistent with undamped ROs in the ECL \cite{tromb,cohen,schunk,helms,RitterJOSA,van}.

Between these two LCs for a narrow  range of $\eta\!\sim\!0.365$ 
is a QP region shown in Fig.\! \ref{per_all}(a2), (b2), (c2), and (d2).
Here, the TSs show a roughly periodic modulation between the two 
LC-like behaviors as seen in Fig.\! \ref{per_all}(b2). The RF spectra,
however, are quite broad as the TSs involve quite abrupt modulation
between the two LCs and involve the participation of ECMs other than
the most dominant ones. This is shown in Fig.\! \ref{per_all}(c2); here, two 
optical frequencies close to $f_{RO}$ dominate, while a broad but 
significant pedestal of other frequencies also plays a role.
The phase portrait is shown in Fig.\! \ref{per_all}(d2).  The gross features
can be reproduced by overlaying the phase portraits Fig.\! \ref{per_all}(d1) and (d3)
of the flanking LCs; however, one notes additional less-pronounced features as well
that arise in part from transients during switching between the two LC-like behaviors.

\subsubsection{Subharmonic} 

As $\eta$ is increased into the range 0.4-0.52, the ECL is characterized by SH dynamics 
\cite{MukaiPRL,kao1994IEEE,ahmed2009numerical}. 
By SH we mean that the TSs are dominated by two (or more) frequencies that sum to 
$f_{RO}\pm mf_{\tau}$ with $m$ an integer. 
In selected cases, we shall find it useful to plot phase portraits obtained from relatively short-duration
TSs to differentiate qualitatively different behaviors that contribute to the SH dynamics
at a given $\eta$.  This will enable us to gain insight into what portions of the trajectory 
are associate with which RF frequencies.

\begin{figure}[h]
\vspace{5pt}
\centerline{\includegraphics[width=.6\textwidth]{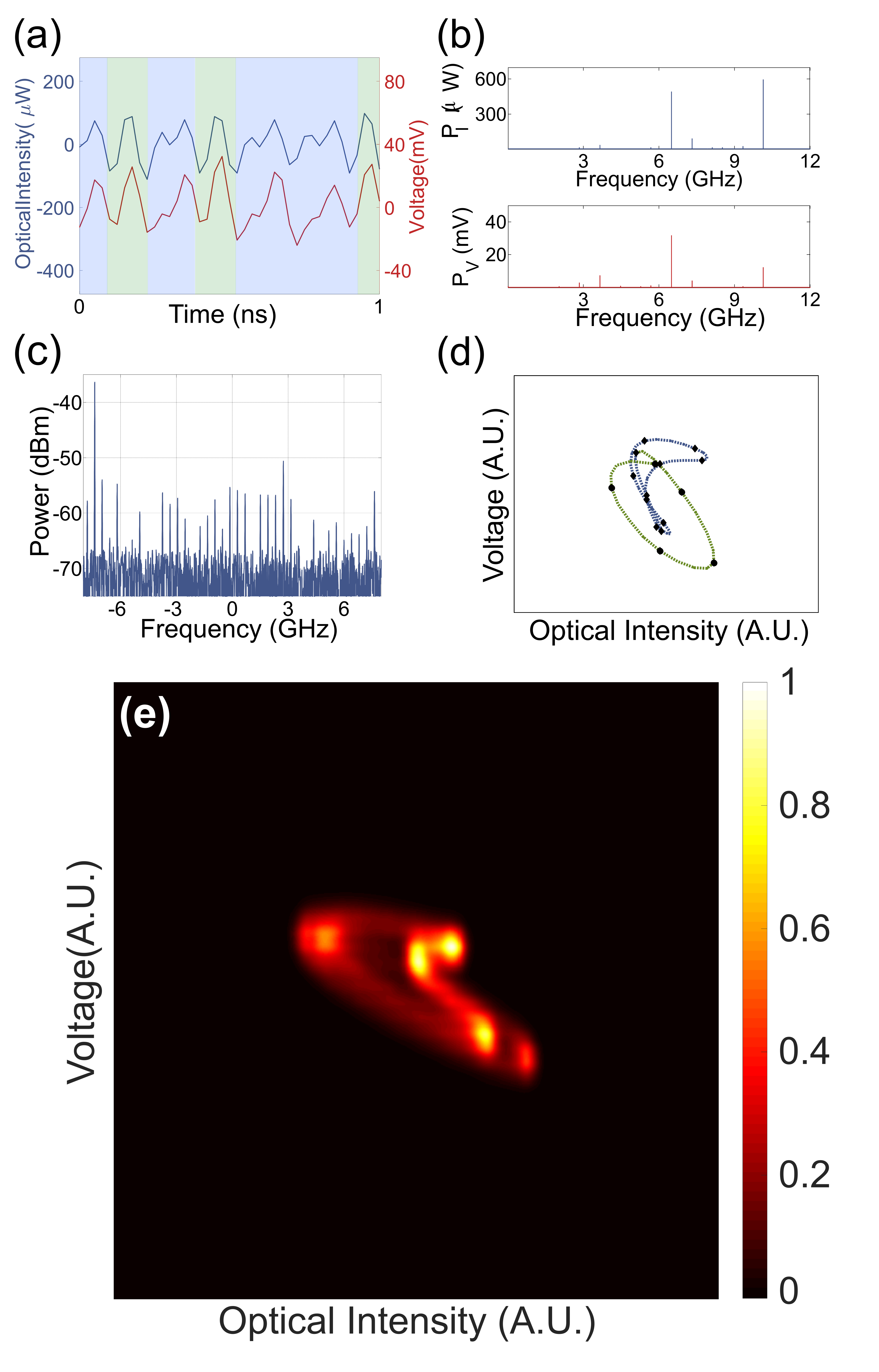}}
\vspace{-5pt}
\caption{A 1 ns zoom-in on the dynamics in the SH regime with $\eta \!= \!0.497$. (a) Shows the TSs and (b) the corresponding RF spectra and (c) the optical spectrum. 
The phase portrait is shown in (d) with short-duration TSs [0.4 ns] and (e) long-duration [25 $\mu$s] TSs.}
\vspace{-10pt}
\label{sh}
\end{figure}

Figure \ref{sh} shows (a) the TSs, (b) the RF spectra, (c) the optical spectra, and the phase portrait (d) with short-duration TSs [0.4 ns] and (e) long-duration [25 $\mu$ second] TSs. The TSs, shown in Fig.\! \ref{sh}(a) exhibit oscillations near $f_{RO}$,
that are themselves modulated at roughly $f_{RO}/2$ indicating
a beating between two frequencies close to $f_{RO}$.
This is supported by the appearance in Fig.\ \ref{sh} (b) of two dominant peaks in the RF spectra with frequencies
$f_1\!=\!3.66$ GHz and $f_2\!=\!6.5$ GHz  (other weaker peaks also appear); note that $f_1\!+\!f_2\!=\!f_{RO}\!= \!10.16$ GHz. 
The phase portrait using relatively short-duration TSs in Fig.\ \ref{sh} (d)
emphasizes the nature of the dynamics. 
In Fig.\ \ref{sh} (e) we see two open structures that  repeat in succession, that can be distinguished as the green and blue portions of the phase portrait in Fig.\! \ref{sh} (d)
and corresponding shaded regions of Fig.\! \ref{sh}(a). 
The RF frequency [Fig.\! \ref{sh}(b)], 
$f_1\!=\!3.66 $ GHz corresponds to the green structure 
and $f_2\!=\!6.5$ GHz corresponds to the blue structure.

\begin{figure}[h]
\vspace{5pt}
\centerline{\includegraphics[width=.6\textwidth]{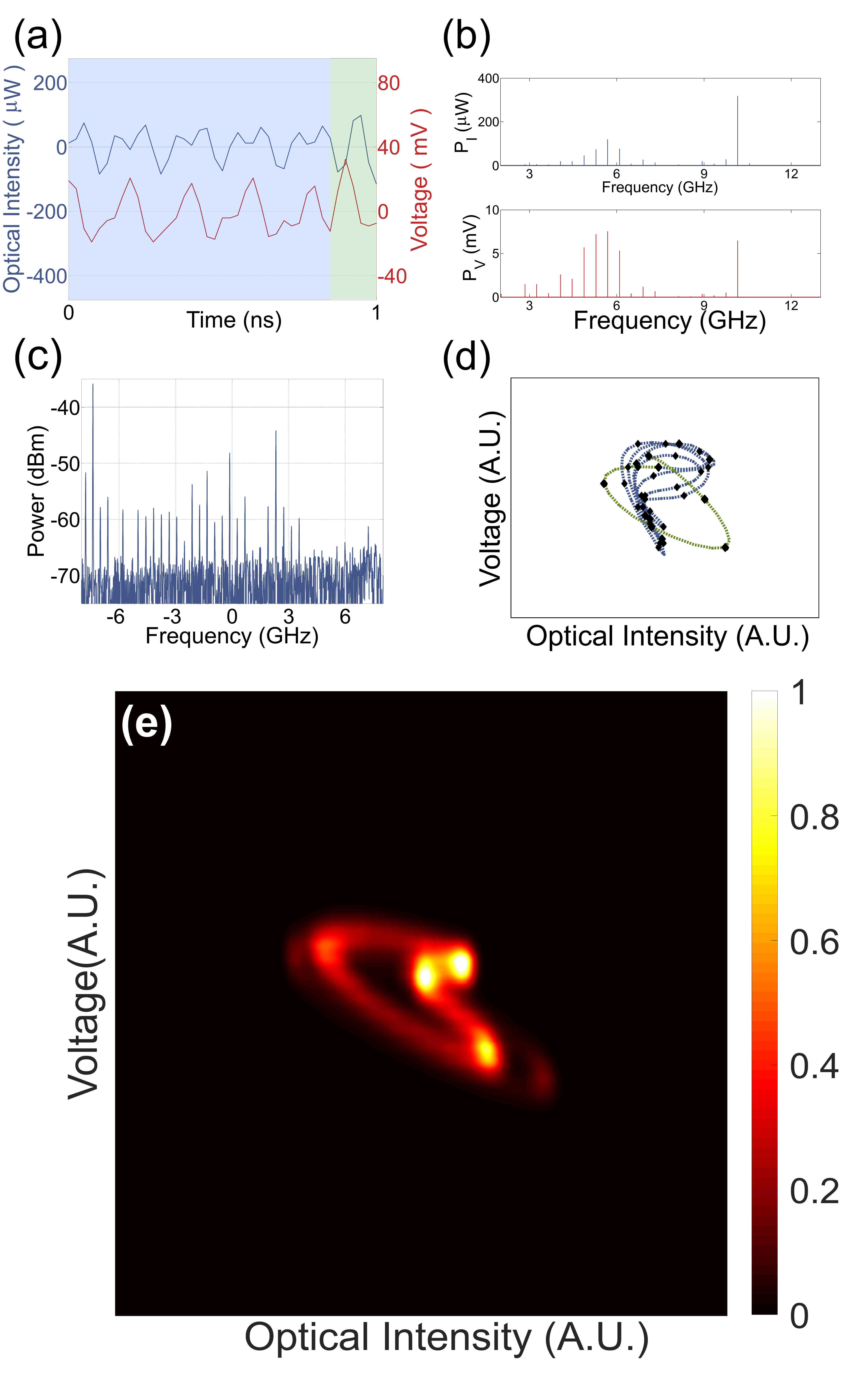}}\vspace{-5pt}
\caption{A 1 ns zoom-in on the dynamics in the different SH regime with $\eta\! =\! 0.517$.
(a) Shows the TSs and (b) the corresponding RF spectra and (c) the optical spectrum. 
The phase portrait is shown in (d) with short-duration TSs [0.4 ns] and (e) long-duration [25 $\mu$s] TSs.}
\vspace{-10pt}
\label{perpd}
\end{figure}

Figure \ref{perpd} shows an example restricting our view to 
a different SH behavior for $\eta\! =\! 0.517$. 
In Fig.\! \ref{perpd} (a) a shorter timescale is shown (1 ns) compared with that used in previous figures. The TSs appear roughly as sinusoidal modulated by an envelope at half the dominant frequency of the sinusoid [more apparent in the TS for $I(t)$].  This is the signature of the PD regime. 
What is evident from the TSs is born out in the RF spectra in Fig.\! \ref{perpd}(b). Here we see a dominant peak at $f_{RO}$ and a cluster of peaks near $f_{RO}/2$. Figures \ref{perpd}(d) and (e) show the corresponding phase portraits, for (d) short-duration TSs [0.4 ns] and (e) long-duration [25 $\mu$s] TSs. Compared with what was seen in the SH regime at somewhat smaller $\eta$, far more time is spent on the 
blue trace and shaded region with (irregularly shaped) features reminiscent of PD dynamics 
than on the green trace and shaded region with features resembling behavior in the LC 
regime discussed above, in Fig.\! 7 (a) and (d). The RF spectrum corresponding to Fig.\! \ref{perpd}(b) shows a 
dominant peak at  $f_{RO} \!= \!10.16$ GHz while the TSs in stable PD 
 have dominant peak at $f_{RO}+f_\tau \!= \!10.56$. 
 A smoothed curve connecting both regimes where the green (blue) curve is determined to be 
LC (PD). 
 The two trajectories are mapped on the phase portrait in Fig.\ \ref{perpd}(d) 
 where the PD regime is plotted in blue; 
 and the LC trajectory is plotted in green. 
 The diamonds (circles) are the raw data from the TSs. 
 Lastly, the corresponding phase portrait for long TSs is shown in Fig.\! \ref{perpd}(e). 

\subsubsection{Period-doubled and intermittency}

\begin{figure}[h]
	\vspace{-1pt}
	\centerline{\includegraphics[width=.6\textwidth]{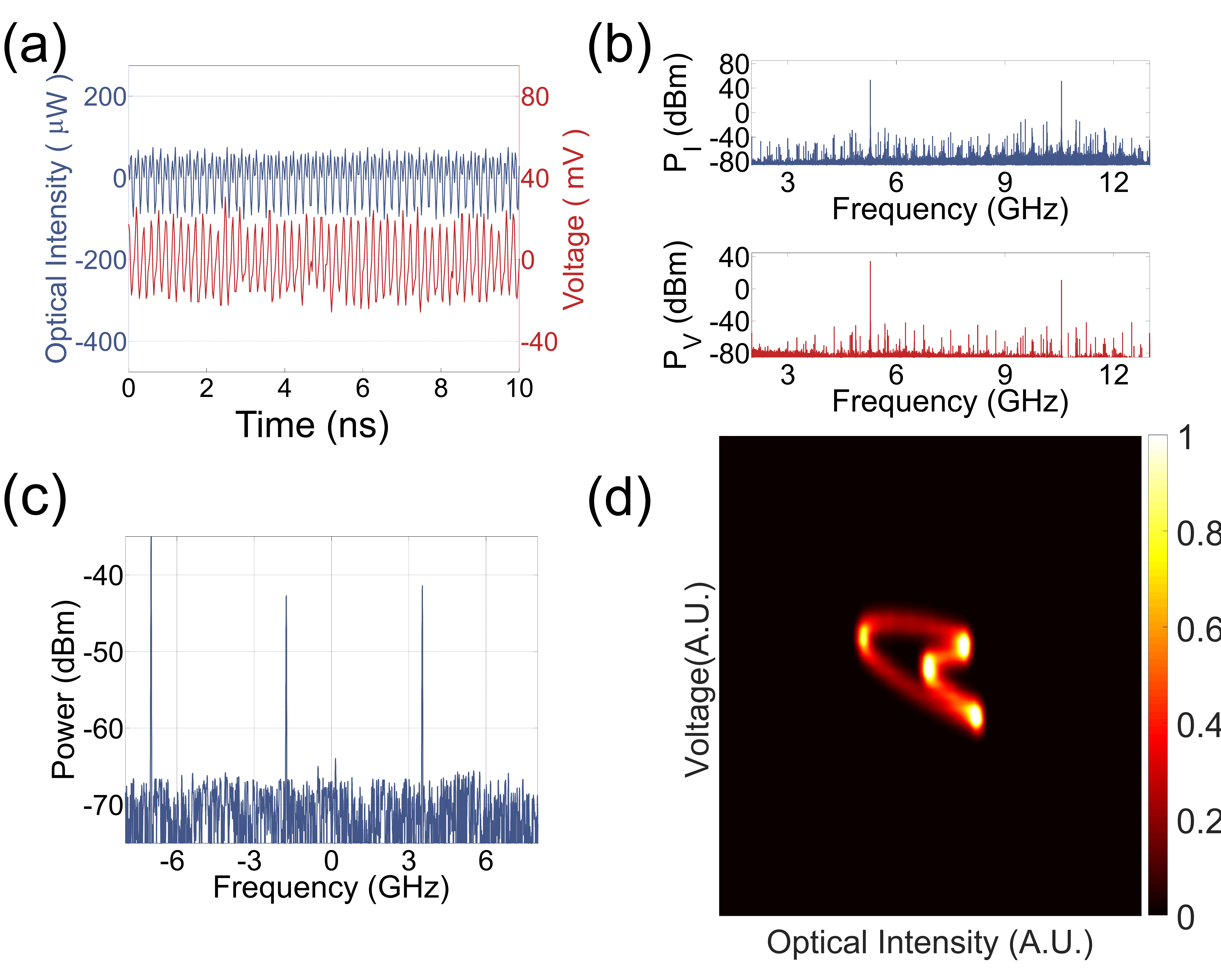}}
	\vspace{-5pt}
	\caption{Dynamics in the PD regime with $\eta\! = \!0.55$.  
		(a) The TSs, (b) RF spectrum, (c) optical spectrum, and (d)  phase portrait.}
	\label{pd}
\end{figure}

 The PD regime for $\eta\!=\!0.52$ to 0.64 is characterized by RF 
 frequencies  $f_{RO}\! = \!10.16$ GHz and
 $f_{RO}/2$. We show here an example for $\eta\!=\!0.55$, where
 we observe a PD regime as plotted in Fig.\! \ref{pd}, with the TSs  in (a), the RF spectra in (b),
 the optical spectrum in (c), and the phase portrait in (d). 
 The main features are similar to those in Fig.\! \ref{perpd}.
 The dominant RF frequencies  in Fig.\ \ref{pd}(b) are 
 $f_{RO}\!+\!f_\tau = \!10.56$ GHz and 
 $(f_{RO}\!+\!f_\tau)/2\! = \! 5.28$ GHz 
 compared with $f_{RO}\!=\!10.16$ GHz and $f_{RO}/2\!=\!5.08$ GHz   in Fig.\! \ref{perpd}. 
 Note that even though the phase portrait shown in Fig.\ \ref{pd}(d) 
 has similar overall shape as the 
 phase portrait (green) in Fig. \ref{perpd}(b), 
 the two TSs do not have the same dominant  RF frequencies. 
The transition of the dynamical behavior varies from SH to PD involves the formation of a stable solution. The RF peak in SH is found to be $f_{\tau}$ + $f_{RO}$ and the trajectories are comprised of an LC and a PD, as the feedback strength increases. As feedback is increased, we find the duration of PD increases, and the trajectories become more dominant until PD takes over and the new solution is found to be PD with RF peak at $f_{RO}$.
  
We illustrate intermittency at $\eta \! = \!0.545$ which is evident within a single TS shown in Fig.\ \ref{wch} and the animation in the supplement material. This type of intermittency has been investigated in Ref. \cite{RiderPRE}, we can identify chaotic and PD along with other complex dynamics shown in Figs.\!  \ref{wch}(a)-(d). The sequential order of the dynamics in this TS appear to be random which can be found in the animation in the supplement material. We see considerable broadening of the features compared with the PD phase portrait in Fig.\! \ref{perpd}. The phase portrait exhibits an overall elliptical structure 
 (LC-like behavior) together with enhanced density in the upper right of the phase  portrait associated with PD-like behavior, with considerably more time spent on the  PD-like phase ($\sim\!35\!:\!1$). A more detailed examination of the phase portrait over many shorter duration TSs bears out this interpretation. The PD-like dynamics here differ somewhat from those shown in Fig.\! \ref{pd}.

\begin{figure}[h]
	\vspace{-1pt}
	\centerline{\includegraphics[width=.6\textwidth]{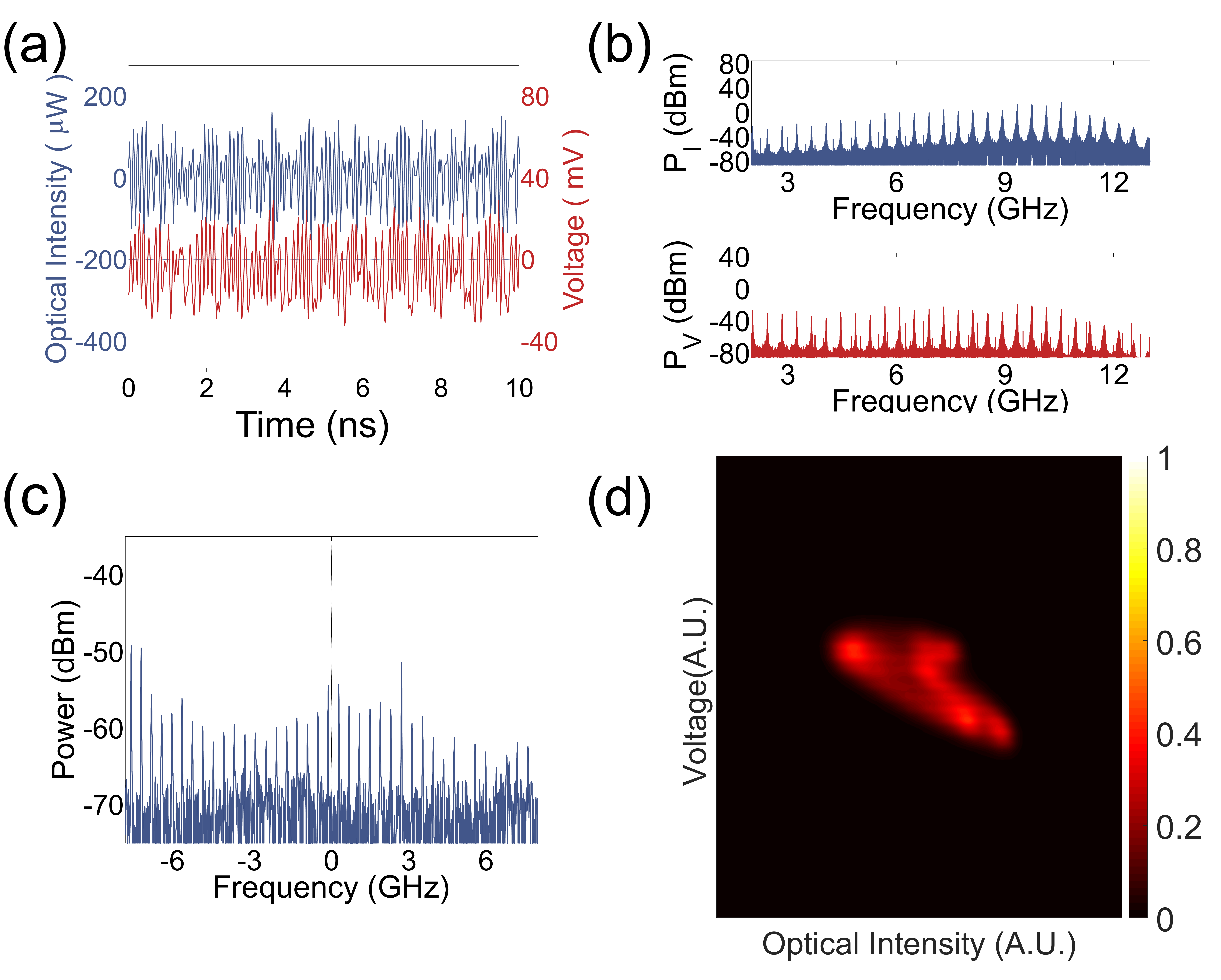}}
	\vspace{-5pt}
	\caption{Dynamics in the intermittent regime with $\eta\! =\! 0.55$. 
		(a) The TSs, (b) RF spectra,  (c) optical spectrum, and 
		(d)  phase portrait.}
	\vspace{-5pt}
	\label{wch}
\end{figure}

\subsubsection{Coherence collapse}

\begin{figure}[h]
\vspace{-8pt}
\centerline{\includegraphics[width=.6\textwidth]{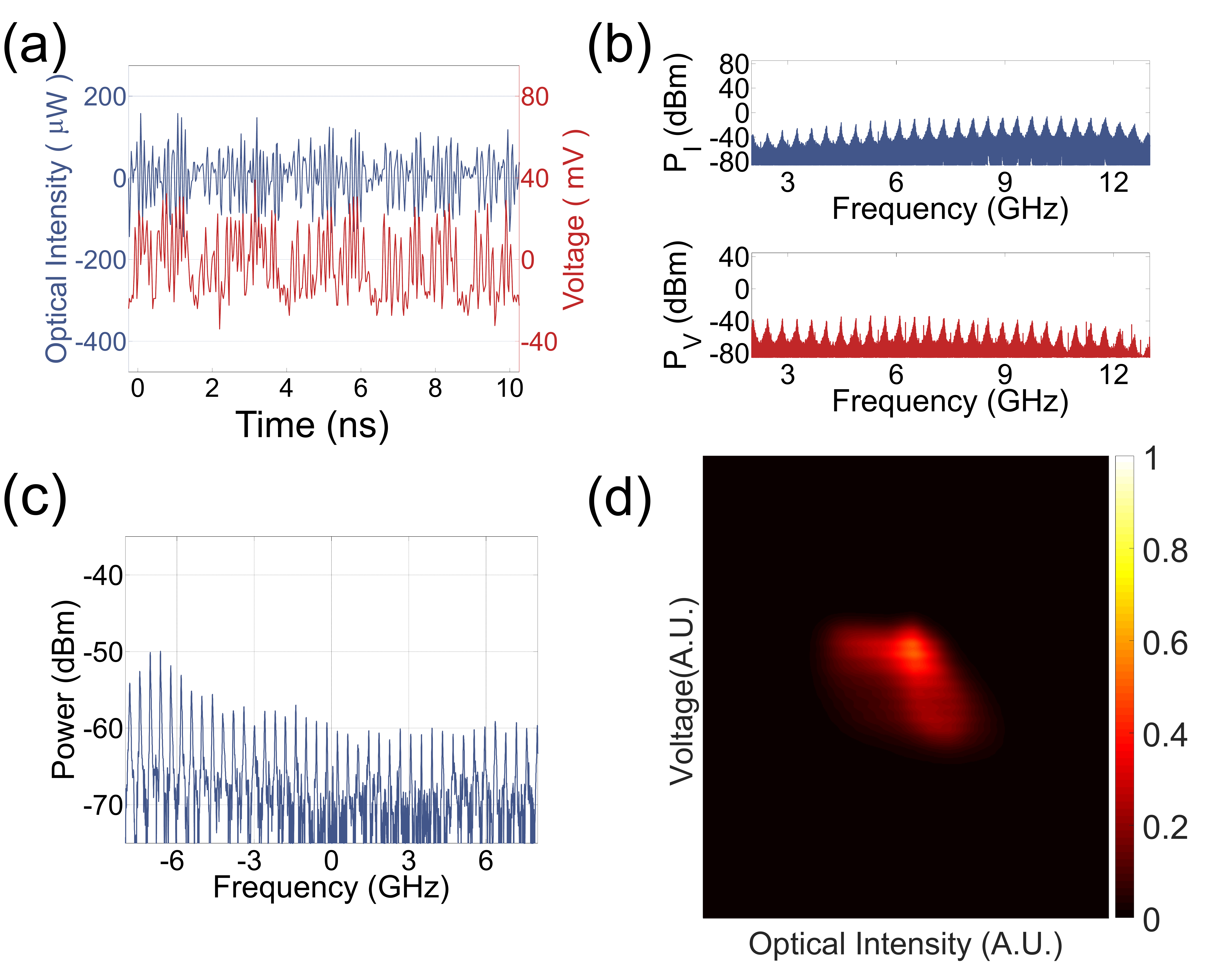}}
\vspace{-5pt}
\caption{Dynamics in CC regime with $\eta\! = \!0.75$ showing
(a) the TSs, (b) RF spectra, (c) optical spectrum, and (d)  phase portrait.}
\vspace{-5pt}
\label{chaos}
\end{figure}

As $\eta$ is further increased above 0.64 
the onset of chaos is reached. An example for $\eta\!=\!0.75$ is shown in Fig.\! \ref{chaos}.
In this regime, the BDs of Fig.\! \ref{bd2}(a) and (b) are quite broadened in the vertical 
direction indicating many maxima and minima  in the respective TSs
as well as the participation of numerous ECMs [Fig.\! \ref{bd2}(c)].
The TSs [Fig.\! \ref{chaos}(a)] become quite complex and involve a very broad range of RF frequencies
 [Fig.\! \ref{chaos}(b)].  Likewise, many ECMs participate in the dynamics [Fig.\! \ref{chaos}(c)]. 
The phase portrait [Fig.\! \ref{chaos}(d)], consequently, has an undifferentiated broadened 
feature with a hint of elliptical structure (reflecting the continues role of 
RO oscillations).


\section{Conclusion}

The complex dynamics of ECLs correspond to trajectories in phase space
whose qualitative nature undergoes pronounced changes as a control parameter
(here $\eta$) is varied.  
Until now, the overwhelming majority of experimental results have been obtained 
on just one dynamical degree of freedom, \textit{viz.}\ $I(t)$. 
By simultaneously measuring $I(t)$ and $V(t)$, however, we obtain new insight into the 
nonlinear dynamics of ECLs based on two degrees of freedom; 
in our case projecting  phase space in the $\phi$ direction. 
The resulting phase portraits can be used to track the trajectory's motion 
between the various ECMs of the ECL. 
The importance of the ability to visualize the dynamics itself should not be 
underestimated.  As shown here, that ability enables us to attain 
insight into rather complex dynamical regimes that would be otherwise difficult 
to understand.
Moreover, as our results illuminate two degrees of freedom, 
they show promise to more thoroughly test the LK model than can be provided by a 
single dynamical variable  \cite{DaanPR,wieczorek2003bifurcation}.   
In the work presented here, 
we observe a QP route to chaos interrupted by two types of LCs, \textit{viz.}\ a SH regime and a coexisting PD before entering the edge of CC.  

 We gratefully acknowledge the financial support of the Conseil Regional de Lorraine. The authors gratefully acknowledge the financial support of the 2015 Technologies Incubation scholarship from Taiwan Ministry of Education for Mr. C. Y. Chang.



\begin{thebibliography}{9}
\bibitem{note1}
Ref.\ \cite{BrunnerPRL} maps all three dynamical variables simultaneously in the LFF
regime for an ECL. In our work, we consider more broadly the route to chaos in an 
ECL.  Moreover, the route to chaos in our ECL is a different route (commented upon in the body of
our paper) and does not involve LFF.
\end{thebibliography}
\end{document}